\documentstyle[eqsecnum,pre,preprint,aps,epsfig]{revtex}

\tightenlines

\newcommand{\bq}{\begin{equation}}
\newcommand{\eq}{\end{equation}}
\newcommand{\bqa}{\begin{eqnarray}}
\newcommand{\eqa}{\end{eqnarray}}
\newcommand{\ben}{\begin{enumerate}}
\newcommand{\een}{\end{enumerate}}
\newcommand{\bc}{\begin{center}}
\newcommand{\ec}{\end{center}}

\def\gb{\bar{\mathnormal g}}

\def\etal{{\it et.al.\/}}

\global\nulldelimiterspace = 0pt

\def\L{ {\cal L }}

\def\A{ {\cal A }}

\begin{document}

\draft
\preprint{PM/99-45, corrected version}

\title{$Z$-peak subtracted representation of Bhabha
scattering and search for new physics
effects}

\author{M. Beccaria$^{a,b}$ F.M.
Renard$^c$, 
S. Spagnolo$^d$ and C. Verzegnassi$^{e,f}$ \\
\vspace{0.4cm} 
}

\address{
$^a$Dipartimento di Fisica, Universit\`a di 
Lecce \\
Via Arnesano, 73100 Lecce, Italy.\\
\vspace{0.2cm}  
$^b$INFN, Sezione di Lecce\\
Via Arnesano, 73100 Lecce, Italy.\\
\vspace{0.2cm} 
$^c$ Physique
Math\'{e}matique et Th\'{e}orique, UMR 5825\\
Universit\'{e} Montpellier
II,  F-34095 Montpellier Cedex 5.\hspace{2.2cm}\\
\vspace{0.2cm} 
$^d$ 
Rutherford Appleton Laboratory - Particle Physics Department \\
Chilton, Didcot, Oxfordshire OX11 0QX\\
\vspace{0.2cm}
$^e$
Dipartimento di Fisica Teorica, Universit\`a di Trieste, \\
Strada Costiera
 14, Miramare (Trieste) \\
\vspace{0.2cm} 
$^f$ INFN, Sezione di Trieste\\
}

\maketitle

\begin{abstract}
We extend to the special case of Bhabha scattering the "$Z$-peak
subtracted" representation previously applied to $e^+e^-\to f\bar f$
($f\neq e$). This allows us to analyze the process at any energy while
imposing in an automatic way the constraints set by high precision
measurements at the $Z$ peak. The procedure turns out to be
particularly convenient for the search of a certain class of new
physics effects at variable energy. A few examples are considered 
and the information thus obtained is
combined with the corresponding one that can be derived 
from the other $e^+e^-$
annihilation processes, both at present (LEP2) and at future colliders.
This shows that the role of Bhabha scattering in this
respect can be quite relevant.
\end{abstract}


\section{Introduction.} 

A convenient way of searching for virtual new physics effects in
$e^+e^-$ annihilation processes has been recently proposed; it
consists of the
use of the so-called "Z-peak subtracted representation" that allows to
take automatically into account the severe constraints imposed by the
high precision measurements performed at the $Z$ peak by LEP1 and
SLC\cite{LEPcombi}.
This is achieved by choosing as "theoretical inputs" the measured values
of the partial $Z$ widths $\Gamma_{f}$ and of the effective weak angle
$sin^2\theta_{\rm eff}$ together with $\alpha(0)$. For each $e^+e^-\to
f\bar f$  ($f\neq e$) annihilation process, all the one-loop standard
model (SM) or new physics (NP) effects are described by four functions 
of the energy ($\sqrt{q^2}$) and of the scattering angle $\theta$ which are
subtracted at $q^2=M^2_Z$ or at $q^2=0$ in order to take
the inputs into account. We defer to ref.\cite{Zsub} for
further details.\par
This procedure is especially suitable for the search of NP
effects which grow with the energy. Various applications were made for
Supersymmetry , Technicolour , anomalous gauge
couplings, higher $Z'$ bosons,  4-fermion contact terms
and extra-dimensions \cite{Zsub,applic,agczsub,tc,deltai}. 
The description was shown \cite{deltai} to be 
particularly useful when NP is characterized by an effective scale
$\Lambda$ which is much higher than the actual energy range of
$\sqrt{q^2}$. In the particular case of \underline{universal} 
$\theta$ independent  effects
all the information on NP can be then conveyed into three constants
called, $\delta_{Z,s,\gamma}$, that can be viewed as the generalization,
beyond the $Z$ peak, of the $T$, $S$ \cite{Peskin}or $\epsilon_{1,3}$
\cite{AB} description.\par
This representation was not yet applied to Bhabha scattering because
of the complication generated by the presence of $t$-channel photon
and $Z$ exchanges. 
The purpose of the present paper is to fill this
lack. We shall show here in Section 2
that the whole $Z$-peak subtracted formalism can 
be extended  to Bhabha scattering in a very natural way. This 
will allow to use this process, together with the
other $e^+e^-\to f\bar f$ ones, in order to improve the constraints on
possible NP contributions, as illustrated in Section 3. 
In Section 4, we shall show in the numerical  applications that
the gain thus obtained can be substancial.

\section{The $Z$-peak subtracted representation}

We shall first summarize the results of the $Z$-peak
subtracted representation, described in previous dedicated 
papers \cite{Zsub}, by writing the general $e^+e^-\to f\bar f$
($f\neq e$) scattering amplitude at one loop as the sum of an
effective photon and an effective $Z$ amplitude with couplings
$g^{\gamma}_{Vj}(q^2,\theta)$, $g^{Z}_{Vj}(q^2,\theta)$, 
$g^{Z}_{Aj}(q^2,\theta)$, where the index $j$ denotes either the
initial electron ($j=e$) or the final fermion ($j=f\neq e$).

\bqa
&&\A(q^2,\theta)= {i\over q^2}
\bar v(e^+)\gamma^{\mu}g^{(\gamma)}_{Ve}(q^2,\theta)u(e^-).
\bar u(f)\gamma_{\mu}g^{(\gamma)}_{Vf}(q^2,\theta)v(\bar f)
+{i\over q^2-M^2_Z+iM_Z\Gamma_Z}.\nonumber\\
&&
\bar v(e^+)\gamma^{\mu}
[g^{(Z)}_{Ve}(q^2,\theta)-g^{(Z)}_{Ae}(q^2,\theta)\gamma^5]u(e^-).
\bar u(f)\gamma_{\mu}
[g^{(Z)}_{Vf}(q^2,\theta)-g^{(Z)}_{Af}(q^2,\theta)\gamma^5]v(\bar f)
\label{Amp}\eqa

The aforementioned inputs are taken into account by imposing that the
total amplitude takes the required value at $q^2=0$ and at $q^2=M_Z^2$.
Following the method of~\cite{Zsub}, this amounts to use a subtraction
procedure which allows to write:

\bqa
&&g^{\gamma}_{Ve}(q^2,\theta)=\sqrt{4\pi\alpha(0)}~Q_e
[1+{1\over2}\tilde\Delta_{\alpha, ef}(q^2,\theta)]\nonumber\\
&&g^{\gamma}_{Vf}(q^2,\theta)=\sqrt{4\pi\alpha(0)}~Q_f
[1+{1\over2}\tilde\Delta_{\alpha, ef}(q^2,\theta)]\nonumber\\
&&g^{\gamma}_{Ae}(q^2,\theta)=g^{\gamma}_{Af}(q^2,\theta)=0\nonumber\\
&&g^{Z}_{Ve}=\gamma_e^{{1\over2}}~I_{3e}~\tilde v_e
[1-{1\over2}R_{ef}(q^2,\theta)-{4\tilde s_e\tilde c_e\over
\tilde v_e}|Q_f|V^{\gamma Z}_{ef}(q^2,\theta)]\nonumber\\
&&g^{Z}_{Vf}(q^2,\theta)=\gamma_f^{{1\over2}}~I_{3f}~\tilde v_f
[1-{1\over2}R_{ef}(q^2,\theta)-{4\tilde s_e\tilde c_e\over
\tilde v_f}|Q_f|V^{Z\gamma}_{ef}(q^2,\theta)]\nonumber\\
&&g^{Z}_{Ae}(q^2,\theta)=\gamma_e^{{1\over2}}~I_{3e}
[1-{1\over2}R_{ef}(q^2,\theta)]\nonumber\\
&&g^{Z}_{Af}(q^2,\theta)=\gamma_f^{{1\over2}}~I_{3f}
[1-{1\over2}R_{ef}(q^2,\theta)]\nonumber\\
\label{geff}\eqa
with the $Z$-peak inputs
\bq
\gamma_j^{{1\over2}}=[{48\pi\Gamma_j\over N_jM_Z(1+\tilde
v^2_j)}]^{{1\over2}}
\label{gamma}\eq
\noindent
and
\bq
\tilde v_j=1-4|Q_j|\tilde s^2_j
\label{vj}\eq
\noindent
where $\tilde s^2_j=1-\tilde c^2_j$ is 
the weak effective angle measured
through the forward-backward or polarization asymmetries in the
final channel $j$, 
$\tilde s_e \equiv \tilde s_\mu \equiv \tilde s_\tau$
and $N_j$
is the colour factor with QCD corrections at $Z$-peak.\par
The quantities $\tilde\Delta_{\alpha, ef}(q^2,\theta)$,
$R_{ef}(q^2,\theta)$,
$V^{\gamma Z}_{ef}(q^2,\theta)$, $V^{Z\gamma}_{ef}(q^2,\theta)$ 
contain all the $q^2,~\theta$ dependent parts
of the scattering amplitude due to SM or NP at one-loop.
In our approach, they consist of certain finite combinations
of self-energies, vertices and boxes that are 
\underline{automatically} gauge independent.

For an additional four fermion amplitude

\bq
\bar v(e^+)\gamma^{\mu}[a(q^2,\theta)-b(q^2,\theta)\gamma^5]u(e^-).
\bar u(f)\gamma_{\mu}[c(q^2,\theta)-d(q^2,\theta)\gamma^5] v(f)
\eq
where $a$, $b$, $c$ and $d$ are ${\cal O}(\alpha)$ quantities, one easily 
obtains the corresponding projections on the photon and $Z$ Lorentz 
structures:

\bqa
&&\tilde\Delta_{\alpha, ef}(q^2,\theta)=
q^2{[a(q^2,\theta)-b(q^2,\theta)\tilde v_e]
[c(q^2,\theta)-d(q^2,\theta)\tilde v_f]
\over e^2 Q_e Q_f}\nonumber\\
&&R_{ef}(q^2,\theta)=-(q^2-M^2_Z){4\tilde{s}^2_e \tilde{c}^2_e 
b(q^2,\theta)d(q^2,\theta)
\over e^2 I_{3e}I_{3f}}\nonumber\\
&&
V^{\gamma Z}_{ef}(q^2,\theta)=-(q^2-M^2_Z)
{[a(q^2,\theta)-b(q^2,\theta)\tilde v_e]2\tilde{s}_e \tilde{c}_e d(q^2,\theta)
\over e^2 Q_eI_{3f}}\nonumber\\
&&
V^{Z\gamma}_{ef}(q^2,\theta)=-(q^2-M^2_Z){[c(q^2,\theta)-d(q^2,\theta)
\tilde v_f]2\tilde{s}_e \tilde{c}_e b(q^2,\theta)
\over e^2 Q_fI_{3e}}
\label{DRV}\eqa

We have given in ref.\cite{Zsub} and
in the Appendix of ref.\cite{log} the expression 
of the general polarized $e^+e^-\to f\bar f$ differential cross
section in terms of these
four functions. From this one obtains, for example,
the integrated cross sections
$\sigma_f$ and the asymmetries $A_{FB,f}$, $A_{LR,f}$.\par

To generalize our approach to the case $f=e$, Bhabha scattering, it is
convenient to write the scattering amplitude at one loop 
as the sum of two (s-channel and t-channel) components:
\bq
{\cal A}_{ee}={\cal A}_s(q^2, \theta)+{\cal A}_t(q^2, \theta)
\eq

The procedure that we have illustrated 
in the $e^+e^-\to\bar{f}f$ ($f\neq e$) case
applies directly 
to the $s$-channel part of the Bhabha
amplitude. In this case we can
drop the index $ef$, as $e\equiv f$
and we have
$V^{\gamma Z}(q^2,\theta)\equiv V^{Z\gamma}(q^2,\theta) \equiv
 V(q^2,\theta)$, so that we only deal with three independent functions 
$\tilde\Delta_{\alpha}(q^2,\theta)$, $R(q^2,\theta)$ and
$V(q^2,\theta)$.\par
It is now straightforward to check that the same
procedure can be applied step by step
to the $t$-channel component.
In full generality the latter can be written as

\bqa
&&\A_t(q^2, \theta)= {i\over t}
\bar v(e^+)\gamma^{\mu}\bar g^{(\gamma)}_{Ve}(q^2,\theta)v(e^+).
\bar u(e^-)\gamma_{\mu}\bar g^{(\gamma)}_{Vf}(q^2,\theta)u(e^-)
+{i\over t-M^2_Z}.\nonumber\\
&&
\bar v(e^+)\gamma^{\mu}
[\gb^{(Z)}_{Ve}(q^2,\theta)
-\gb^{(Z)}_{Ae}(q^2,\theta)\gamma^5]v(e^+)
\bar  u(e^-)\gamma_{\mu}
[\gb^{(Z)}_{Vf}(q^2,\theta)
-\gb^{(Z)}_{Af}(q^2,\theta)\gamma^5]u(e^-)\nonumber\\
\label{Ampt}\eqa
\noindent
with the $t$-channel effective couplings:

\bqa
&&\gb^{\gamma}_{Ve}(q^2,\theta)
=\sqrt{4\pi\alpha(0)}~Q_e
[1+{1\over2}\overline {\tilde\Delta}_{\alpha}(q^2,\theta)]\nonumber\\
&&\gb^{Z}_{Ve}(q^2,\theta)=\gamma_e^{{1\over2}}~I_{3e}~\tilde v_e
[1-{1\over2}\overline R(q^2,\theta)-{4\tilde s_e\tilde c_e\over
\tilde v_e}|Q_f|\overline V(q^2,\theta)]\nonumber\\
&&\gb^{Z}_{Ae}(q^2,\theta)=\gamma_e^{{1\over2}}~I_{3e}
[1-{1\over2}\overline R(q^2,\theta)]\nonumber\\
\label{gefft}\eqa
\noindent
where the new functions $\overline{\tilde\Delta}_{\alpha}(q^2,\theta)$,
$\overline R(q^2,\theta)$ and $\overline V(q^2,\theta)$ 
are obtained from the
previous $s$-channel ones (without bar) by 
($q^2 \longleftrightarrow t$) crossing relations:

\bq
\overline{\tilde\Delta}_{\alpha}(q^2,\theta)=
[{\tilde\Delta}_{\alpha}(q^2,\theta)]( 
q^2 \longrightarrow t=-{q^2\over2}(1-cos\theta);
~cos\theta \longrightarrow 1+{2q^2\over t})
\label{cross}\eq
\noindent
and analogously for $\overline R$, $\overline V$.

The general expression of the polarized Bhabha 
differential cross section obtained
from the sum  of the $s$-channel (\ref{Amp}) and $t$-channel
(\ref{Ampt}) amplitudes is given in the Appendix, in the form:
\bq
{d\sigma\over dcos\theta}=(1-PP'){d\sigma^{1}\over dcos\theta}
+(1+PP'){d\sigma^{2}\over dcos\theta}
+(P'-P){d\sigma^{P}\over dcos\theta}
\eq
where $P$ and $P'$ are the initial $e^-$, $e^+$ polarizations.

We shall write the three differential cross sections as the sum of
a Born term and a one loop contribution, $d\sigma^i = (d\sigma^i)^{Born} +
(d\sigma^i)^{(1)}$.

The Born term is given by

\bqa
&&({d\sigma^{1}\over dcos\theta})^{Born}={\pi\alpha^2\over q^2}\{
{t^2+u^2\over q^4}+{u^2\over t^2}+{2u^2\over tq^2}\nonumber\\
&&+2({3\Gamma_e\over\alpha M_Z})[{(u^2-t^2+\tilde
v^2_e(u^2+t^2))(q^2-M^2_Z)\over(1+\tilde v^2_e)
q^2[(q^2-M^2_Z)^2+M^2_Z\Gamma^2_Z]}+{u^2\over q^2(t-M^2_Z)}+
{u^2(q^2-M^2_Z)\over t[(q^2-M^2_Z)^2+M^2_Z\Gamma^2_Z]}\nonumber\\
&&+
{u^2\over t(t-M^2_Z)}]
+({3\Gamma_e\over\alpha M_Z})^2[{(t^2+u^2)(1+\tilde v^2_e)^2+
4\tilde v^2_e(u^2-t^2)\over(1+\tilde v^2_e)^2
[(q^2-M^2_Z)^2+M^2_Z\Gamma^2_Z]}
\nonumber\\
&&+{u^2[(1+\tilde v^2_e)^2
+4\tilde v^2_e]\over(1+\tilde
v^2_e)^2(t-M^2_Z)}({2(q^2-M^2_Z)\over
[(q^2-M^2_Z)^2+M^2_Z\Gamma^2_Z]}+{1\over(t-M^2_Z)})]\}   
\label{BN1}\eqa

\bqa
&&({d\sigma^{2}\over dcos\theta})^{Born}={\pi\alpha^2q^2}\{{1\over
t^2}+2({3\Gamma_e\over\alpha M_Z}){(\tilde v^2_e-1)\over(1+\tilde
v^2_e)t(t-M^2_Z)}\nonumber\\
&&+({3\Gamma_e\over\alpha M_Z})^2
{(1-\tilde v^2_e)^2\over(1+\tilde
v^2_e)^2(t-M^2_Z)^2}\} 
\label{BN2}\eqa

\bqa
&&({d\sigma^{P}\over dcos\theta})^{Born}={4\pi\alpha^2u^2\over q^2}
[{\tilde v_e\over(1+\tilde v^2_e)}]\{
({3\Gamma_e\over\alpha M_Z})({1\over
q^2}+{1\over t})
[{(q^2-M^2_Z)\over [(q^2-M^2_Z)^2+M^2_Z\Gamma^2_Z]}
+{1\over(t-M^2_Z)}]\nonumber\\
&&+({3\Gamma_e\over\alpha M_Z})^2[
{1\over
[(q^2-M^2_Z)^2+M^2_Z\Gamma^2_Z]}(1+{2(q^2-M^2_Z)\over(t-M^2_Z)}) 
+{1\over(t-M^2_Z)^2}]\}
\label{BP}\eqa

\noindent
and the one loop contributions of the three functions can be written
in the condensed way:

\bqa
&&({d\sigma^{1}\over dcos\theta})^{(1)}={\pi\alpha^2\over q^2}\{
(t^2+u^2)G_1(q^2,q^2)+(u^2-t^2)G_2(q^2,q^2)\nonumber\\
&&+u^2[G_1(t,t)+G_2(t,t)
+2G_1(q^2,t)+2G_2(q^2,t)]\}
\label{NPN1}\eqa

\bqa
&&({d\sigma^{2}\over dcos\theta})^{(1)}={\pi\alpha^2q^2}
[G_1(t,t)-G_2(t,t)]
\label{NPN2}\eqa

\bqa
&&({d\sigma^{P}\over dcos\theta})^{(1)}={4\pi\alpha^2u^2\over q^2}[
G_3(q^2,q^2)+G_3(q^2,t)+G_3(t,q^2)+G_3(t,t)]
\label{NPP}\eqa

with

\bqa
&&G_1(x,y)={\tilde\Delta_{\alpha}(x)+
{\tilde\Delta}_{\alpha}(y)\over
xy}\nonumber\\
&&+({3\Gamma_e\over\alpha M_Z})
{\tilde v^2_e\over(1+\tilde
v^2_e)}[~{\tilde\Delta_{\alpha}(x)
-R(y)-
{8\tilde s_e\tilde c_e\over\tilde v_e}V(y)\over
x(y-M^2_Z)}+{\tilde\Delta_{\alpha}(y)- R(x)-
{8\tilde s_e\tilde c_e\over\tilde v_e} V(x)~\over
y(x-M^2_Z)}]\nonumber\\
&&-({3\Gamma_e\over\alpha M_Z})^2[~{R(x)+R(y)
+{8\tilde s_e\tilde c_e\tilde v_e\over(1+\tilde v^2_e)}(V(x)+
V(y))~\over(x-M^2_Z)(y-M^2_Z)}]\}
\label{G1}\eqa

\bqa
&&G_2(x,y)=({3\Gamma_e\over\alpha M_Z})
{1\over(1+\tilde
v^2_e)}[~{\tilde\Delta_{\alpha}(x)-R(y)\over
x(y-M^2_Z)}+{{\tilde\Delta}_{\alpha}(y)- R(x)~\over
y(x-M^2_Z)}]\nonumber\\
&&-({3\Gamma_e\over\alpha M_Z})^2{4\tilde v^2_e\over(1+\tilde
v^2_e)^2}[~{R(x)+R(y)
+{4\tilde s_e\tilde c_e\over\tilde v_e}(V(x)+
 V(y))~\over(x-M^2_Z)(y-M^2_Z)}]\}
\label{G2}\eqa

\bqa
&&G_3(x,y)=
{\tilde v_e\over(1+\tilde
v^2_e)}
\{({3\Gamma_e\over\alpha M_Z})
~{\tilde\Delta_{\alpha}(x)-R(y)-
{4\tilde s_e\tilde c_e\over\tilde v_e}V(y)\over
x(y-M^2_Z)}\nonumber\\
&&-({3\Gamma_e\over\alpha M_Z})^2 [~{R(x)+R(y)
+{4\tilde s_e\tilde c_e\over\tilde v_e}V(x)
+{8\tilde s_e\tilde c_e\tilde v_e\over
(1+\tilde v^2_e)}
V(y)~\over(x-M^2_Z)(y-M^2_Z)}]\}
\label{G3}\eqa
\noindent
where we use a condensed notation
(for $x,~y$ corresponding to $q^2$ or $t$), 
$\tilde\Delta_{\alpha}(q^2)$ meaning
$\tilde\Delta_{\alpha}(q^2,\theta)$ and $\tilde\Delta_{\alpha}(t)$
meaning $\overline{\tilde\Delta_{\alpha}}(q^2,\theta)$; similarly for
$R$ and $V$.\par

>From the three previous quantities $d\sigma^{1,2,P}$ we can compute
for instance the unpolarized angular 
distribution\\
 \bq
{d\sigma\over dcos\theta}\equiv{d\sigma^1\over dcos\theta}+
{d\sigma^2\over dcos\theta}
 \eq
the Left-Right polarization asymmetry
 \bq
A_{LR}(q^2, \theta) = 
[{d\sigma^P\over dcos\theta}]/[{d\sigma\over dcos\theta}]
 \eq
and the new (LL+RR)/(LR+RL+LL+RR) polarization asymmetry which arises
from the typical $t$-channel scattering amplitude
 \bq
A_{||}(q^2, \theta)=[{d\sigma^2\over dcos\theta}]/[{d\sigma\over
dcos\theta}]
 \eq

\section{Applications to several NP models} 

\subsection{Universal NP with a high scale }

The previous representation eqs.(\ref{NPN1})-(\ref{G3}) continues
to be valid in the presence of new physics (NP) that does not add extra
Lorentz structures to those of the SM. In this case, one simply
decomposes the three general one-loop functions as:

\bq
\tilde\Delta_{\alpha},~R,~V = (\tilde\Delta_{\alpha},~R,~V)^{SM}+
(\tilde\Delta_{\alpha},~R,~V)^{NP}
\eq
\noindent
and compute the (NP) effects on the various observables, once their
contribution to $(\tilde\Delta_{\alpha},~R,~V)$ is specified.\par
For a model of new physics that does not satisfy special simplicity
requests, the calculation of virtual effects in the Bhabha scattering
is affected by a proliferation of terms with respect to the
annihilation process $e^+e^-\to f\bar f$,
($f\neq e$), as one sees immediately from inspection of 
eqs.(\ref{NPN1})-(\ref{G3}). In fact, after $\theta$-integration, one
will find in general a set of different functions of $q^2$ that
correspond to each power of $\theta$ in the integrand. Each set arises
from the \underline{six} original functions
$\tilde\Delta_{\alpha},~\overline{\tilde\Delta}_{\alpha},~R,~\overline
R,~V,~\overline V$, which means to double the corresponding number of
the case $f\neq e$. Although this can be a purely computational
problem, it obviously complicates 
the practical treatment for this process.\par
The situation shows a drastic change for those models of new physics
that satisfy the requests of being, at the same time,
universal, independent of the $s$, $t$ channels
scattering angle (e.g. only 
contributing self-energies and/or vertices), and endowed with 
an intrinsic scale
$\Lambda$ "sufficiently" larger than $\sqrt{q^2}$. 
In fact, in the $Z$-peak subtracted approach, one has by construction
\bq
\tilde\Delta_{\alpha}(0, \theta)= \overline{\tilde\Delta}_{\alpha}(0, \theta)=
R(M^2_Z, \theta)=\overline R(M^2_Z, \theta)= 
V(M^2_Z, \theta)=\overline V(M^2_Z, \theta)=0
\eq
\noindent
For Universal New Physics effects of the previous considered type
one can then write the following parametrization: 

\bq
R^{UNP}(z)={(z-M^2_Z)\over M^2_Z}[~\delta_Z]
\label{RUNP}
\eq
\bq
V^{UNP}(z)={(z-M^2_Z)\over M^2_Z}[~\delta_s]
\label{VUNP}
\eq
\bq
\tilde\Delta_{\alpha}^{UNP}(z)={z\over M^2_Z}[~\delta_{\gamma}]
\label{DUNP}
\eq
\noindent
where $z=q^2,t$.\par
The quantities 
$\delta_{Z, s, 
\gamma}$ will be in general unknown functions of $z$. For $\Lambda^2>>z$
we can reasonably assume that the three functions are smooth.
This means that they could be well approximated by the coefficient
of the \underline{lowest} power in a  $q^2$ expansion that is
$\delta_i(0)$ whenever $\delta_i(0)\neq 0$ (this will be the case in 
the two considered examples).
In this case, the \underline{same} three parameters will describe 
the NP effects both on the $s$ and on the $t$ channel observables.
These parameters are also the same that appear, for the chosen models, in
all the remaining proccesses $e^+e^-\to f\bar f$,
($f\neq e$). This fact allows to combine the theoretical analysis of
the two types of processes \underline{without increase of parameters},
thus improving the accuracy of the conclusions that are reached.\par
The NP expression of the functions $G_i(x,y)$ acquires in this case the
simple form:

\bqa
&&G^{UNP}_1(x,y)={1\over M^2_Z}\{\delta_{\gamma}
[{1\over x}+{1\over y}]\nonumber\\
&&+({3\Gamma_e\over\alpha M_Z})
({\tilde v^2_e\over1+\tilde
v^2_e})[\delta_{\gamma}({1\over x-M^2_Z}+{1\over y-M^2_Z})-(\delta_Z+
{8\tilde s_e\tilde c_e\over\tilde v_e}\delta_s)({1\over x}+{1\over y})]
\nonumber\\
&&-({3\Gamma_e\over\alpha M_Z})^2(\delta_Z+
({8\tilde s_e\tilde c_e\tilde v_e\over1+\tilde v^2_e})\delta_s)
[{1\over x-M^2_Z}+{1\over y-M^2_Z}]\}
\label{G1UNP}\eqa

\bqa
&&G^{UNP}_2(x,y)={1\over M^2_Z}\{({3\Gamma_e\over\alpha M_Z})
({1\over1+\tilde
v^2_e})[\delta_{\gamma}[{1\over x-M^2_Z}+{1\over
y-M^2_Z}]-\delta_Z[{1\over x}+{1\over y}]\nonumber\\
&&-({3\Gamma_e\over\alpha M_Z})^2{4\tilde v^2_e
\over(1+\tilde v^2_e)^2}(\delta_Z+
{4\tilde s_e\tilde c_e\over\tilde v_e}\delta_s)
[{1\over x-M^2_Z}+{1\over y-M^2_Z}]\}
\label{G2UNP}\eqa

\bqa
&&G^{UNP}_3(x,y)={1\over M^2_Z}
\ {\tilde v_e\over
1+\tilde v^2_e}
\ \{({3\Gamma_e\over\alpha M_Z}){}[{\delta_{\gamma}\over y-M^2_Z}-{\delta_Z+
{4\tilde s_e\tilde c_e\over\tilde v_e}\delta_s\over x}]
\nonumber\\
&&-({3\Gamma_e\over\alpha M_Z})^2[{\delta_Z+
{4\tilde s_e\tilde c_e\over\tilde v_e}\delta_s\over  y-M^2_Z}+
{\delta_Z+
{8\tilde s_e\tilde c_e\tilde v_e\over(1+\tilde v^2_e)}\delta_s\over
x-M^2_Z}]\}
\label{G3UNP}\eqa

\vspace{1cm}

The three constants $\delta_Z$, $\delta_s$, $\delta_{\gamma}$ depend on
the chosen model and can be easily determined in each separate case. To
show how this procedure works in practice, we shall provide the
expressions of the $\delta_i$ in a couple of specific cases that meet
our simplicity requests. With this aim, we have considered 
the following models:\par

(1) Anomalous Gauge Couplings (AGC)\\
We used 
the framework of Ref.~\cite{agc} in which the effective
Lagrangian is constructed with dimension six operators respecting
$SU(2)\times U(1)$ and
$CP$ invariance. As shown in Ref.~\cite{agczsub}, 
only two parameters ($f_{DW}$
and $f_{DB}$) survive in the Z-peak subtracted approach.
The explicit expression of the UNP contribution to
$\delta_Z$, $\delta_s$ and $\delta_{\gamma}$ are

\bqa
&&\delta_Z=8\pi\alpha({M^2_Z\over\Lambda^2})
({\tilde c^2_e\over \tilde s^2_e}f_{DW}+
{\tilde s^2_e\over \tilde c^2_e}f_{DB})\nonumber\\
&&\delta_s=8\pi\alpha({M^2_Z\over\Lambda^2})({ \tilde c_e\over 
\tilde s_e}f_{DW}-
{\tilde s_e\over \tilde c_e}f_{DB})\nonumber\\
&&\delta_{\gamma}=-8\pi\alpha({M^2_Z\over\Lambda^2})(f_{DW}+
f_{DB})
\eqa

They satisfy the linear constraint:
\bq
\delta_Z -\frac{1-2\tilde s^2_e}{\tilde s_e \tilde c_e} \delta_s
+\delta_{\gamma} = 0 .
\label{AGCrel}\eq

\vspace{1cm}

(2) Technicolour\\
The second considered model was one of Technicolour type 
with two families of
strongly coupled
resonances ($V$ and $A$) \cite{tc}. The typical UNP 
parameters are the two
ratios $F_A/M_A$ and $F_V/M_V$ where $F_{A, V}$ and 
$M_{A, V}$ are the
couplings and the masses (that in this case play the role of the new
physics scale $\Lambda^{TC} >> q^2$) of the {\it lightest} axial 
and vector resonances.
The contribution to $\delta_Z$, $\delta_s$ and $\delta_{\gamma}$ are

\bqa
&&\delta_Z={\pi\alpha\over \tilde s^2_e\tilde c^2_e}[(1-2\tilde s^2_e)^2
{M^2_Z\over M^4_V}F^2_V+
{M^2_Z\over M^4_A}F^2_A]\nonumber\\
&&\delta_s={2\pi\alpha\over \tilde s_e\tilde c_e}
(1-2\tilde s^2_e)(1-2\tilde s^2_e)^2
{M^2_Z\over M^4_V}F^2_V\nonumber\\
&&\delta_{\gamma}=-4\pi\alpha({M^2_Z\over M^4_V})F^2_V
\eqa

Again, we have a linear constraint in the $(\delta_Z, \delta_s,
\delta_{\gamma})$ space:
\bq
\delta_s = -\left(\frac{1-2\tilde s^2_e}{2\tilde s_e\tilde c_e}
\right) \delta_{\gamma} .
\label{TCrel}\eq
and the conditions
\bq
\delta_{Z,s}>0 \ \ \ \ \ \ \delta_{\gamma}<0
\label{TCcons}\eq

\subsection{Non universal examples}

Strictly speaking, our procedure has been motivated by the possibility
of investigating models of a special universal type, for which the
number of parameters to be determined can be suitably reduced.
But there exist interesting models of NP that, although not of universal
type, can be nevertheless described by a very restricted number of
parameters. In these special simple cases 
our $Z$-peak subtracted procedure can be applied, without invoking
any smoothness assumption, using
the more general expressions of eqs.(\ref{NPN1})-(\ref{G3}). In what
follows, we have considered two cases that seem to us particularly
relevant. These are:\par
(3) Contact terms\\
With the idea of
compositeness (but it applies to any 
virtual NP effect with a high intrinsic scale, 
for example higher vector boson exchanges, 
satisfying chirality conservation) the following interaction
\bqa
\L&=&k_{if}{4\pi\over\Lambda^2}\{\eta_{LL}(\bar
\Psi^i_L\gamma^{\mu}\Psi^i_L)(\bar\Psi^f_L\gamma_{\mu}\Psi^f_L)
+\eta_{RR}(\bar
\Psi^i_R\gamma^{\mu}\Psi^i_R)(\bar\Psi^f_R\gamma_{\mu}\Psi^f_R)
\nonumber\\
&&+\eta_{RL}(\bar
\Psi^i_R\gamma^{\mu}\Psi^i_R)(\bar\Psi^f_L\gamma_{\mu}\Psi^f_L)
+\eta_{LR}(\bar
\Psi^i_L\gamma^{\mu}\Psi^i_L)(\bar\Psi^f_R\gamma_{\mu}\Psi^f_R)\}
\label{ct}
\eqa

\noindent
was first introduced in \cite{contact} for any four fermion interaction
$(i\bar i \to f \bar f)$;
$k_{if}={1\over2}$ for $i\equiv f$, $k_{if}=1$ otherwise;
$\Psi_L=(1-\gamma^5)/2~\Psi$, $\Psi_R=(1+\gamma^5)/2~\Psi$; 
$\eta_{ab}$ are phase factors defining the chirality structure of
the interaction. Various applications have been made for pure
chiral cases $(ij)=LL$ or  $RR$ or  $LR$ or $RL$
(keeping only one $\eta_{ij}=\pm 1$), as well as for mixed cases
like $VV$  ($\eta_{LL}=\eta_{RR}=\eta_{RL}=\eta_{LR}=\pm 1$),
$AA$ ($\eta_{LL}=\eta_{RR}=-\eta_{RL}=-\eta_{LR}=\pm 1$),
$VA$ ($\eta_{LL}=-\eta_{RR}=\eta_{RL}=-\eta_{LR}=\pm 1$),
$AV$ ($\eta_{LL}=-\eta_{RR}=-\eta_{RL}=\eta_{LR}=\pm 1$);
see ref.\cite{Schrempp} for a general discussion.\par
In the $Z$-peak subtracted representation, 
the effect of this interaction
on the $e^+e^-\to f\bar f$ ($f\neq e$)
observables is obtained through the following expressions:

\bqa
&&\tilde\Delta_{\alpha, ef}(q^2,\theta)=
({\pi q^2\over
e^2Q_eQ_f\Lambda^2})[\eta_{LL}(1-v_e)(1-v_f)+\eta_{RR}(1+v_e)(1+v_f)
\nonumber\\
&&
~~~~~~~~~~~~~~~~~~~~~~~~~~~~~~~~~~~~~~
+\eta_{RL}(1+v_e)(1-v_f)+\eta_{LR}(1-v_e)(1+v_f)]
\nonumber\\
&&R_{ef}(q^2,\theta)=
-({4\tilde{s}^2_e
\tilde{c}^2_e\pi (q^2-M^2_Z)\over
e^2I_{3e}I_{3f}\Lambda^2})
[\eta_{LL}+\eta_{RR}
-\eta_{RL}-\eta_{LR}]\nonumber\\
&&
V^{\gamma Z}_{ef}(q^2,\theta)=-({2\tilde{s}_e
\tilde{c}_e\pi (q^2-M^2_Z)\over
e^2Q_eI_{3f}\Lambda^2})
[\eta_{LL}(1-v_e)-\eta_{RR}(1+v_e)
+\eta_{RL}(1+v_e)-\eta_{LR}(1-v_e)]\nonumber\\
&&
V^{Z\gamma}_{ef}(q^2,\theta)=-({2\tilde{s}_e
\tilde{c}_e\pi (q^2-M^2_Z)\over
e^2Q_fI_{3e}\Lambda^2})
[\eta_{LL}(1-v_f)-\eta_{RR}(1+v_f)
-\eta_{RL}(1-v_f)+\eta_{LR}(1+v_f)]
\label{DRVct}\eqa

For each choice of chirality structure, like $LL$, $RR$, $LR$, $RL$, 
$VV$,$AA$,$VA$,$AV$, the effects on the differential cross 
section for two fermion production can be described 
by a single parameter $\Lambda$.\par

In the case of Bhabha scattering, the constraint $\eta_{RL}=\eta_{LR}$
applies, so that the above expression can be put in the form
of eq.(\ref{RUNP}-\ref{DUNP}), with:

\bqa
&&\delta_Z=-({16\tilde s^2_e\tilde c^2_e \pi M^2_Z
\over e^2\Lambda^2})[\eta_{LL}+\eta_{RR}-2\eta_{LR}]\nonumber\\
&&\delta_s=-({4\tilde s_e\tilde c_e\pi M^2_Z
\over e^2\Lambda^2})
[\eta_{LL}(1-v_e)-\eta_{RR}(1+v_e)
+2v_e\eta_{LR}]\nonumber\\
&&\delta_{\gamma}=({\pi M^2_Z\over
e^2\Lambda^2})[\eta_{LL}(1-v_e)^2+\eta_{RR}(1+v_e)^2
+2\eta_{LR}(1-v^2_e)]
\eqa

\vspace{1cm}

(4) Extra dimensions \\
Recently, an intense activity
has been developed on possible low energy effects of graviton exchange. 
The following matrix element for the 
4-fermion process $e^+e^-\to \bar{f} f$~\cite{graviton} is predicted:

\bq
{\lambda\over\Lambda^4}[\bar e\gamma^{\mu}e\bar
f\gamma_{\mu}f(p_2-p_1).(p_4-p_3)-\bar e\gamma^{\mu}e\bar f\gamma^{\nu}f
(p_2-p_1)_{\nu}(p_4-p_3)_{\mu}]
\eq

For this model one finds in the case of Bhabha scattering the following
contributions to $\delta_{\gamma, Z, s}$ that, as one sees, are now 
genuine $q^2$, $\theta$ functions:

\bqa
&&
\delta_{\gamma}=({\lambda q^2 M^2_Z\over\Lambda^4}){(\tilde
v^2_e-2cos\theta)\over e^2}\nonumber\\
&&\delta_Z=-
({\lambda q^2 M^2_Z\over\Lambda^4})
({16\tilde s^2_e\tilde c^2_e\over e^2})\nonumber\\
&&\delta_s=
({\lambda q^2 M^2_Z\over\Lambda^4})({4\tilde s_e\tilde c_e\tilde
v_e\over
e^2})
\eqa
\noindent
Illustrations will be given in the next Section with the normalization
$\lambda=\pm1$.
Note that the $q^2$ factor is purely kinematical and a consequence of 
the higher dimension of the interaction Lagrangian. Note also the
presence of the term proportional to $q^2 cos\theta$ 
in the $s$-channel photon coefficient $\delta_{\gamma}$, which gives
a contribution proportional to $t+2q^2$ in the $t$-channel,
according to eq.(\ref{cross}). This contribution will turn out to give
the largest effect through the interference with the standard photon
exchange amplitude.

\vspace{1cm}

Our theoretical description of new physics effects is at this point
concluded. The final Section 4 will be devoted to a detailed numerical
analysis of the information that can be derived on the involved
parameters by the present (LEP2) and future (LC), using both Bhabha
scattering and all the remaining $e^+e^-\to f\bar f$ processes.

\section{Numerical Results}

\subsection{LEP2 (present and future)}

As a first application of our approach, we have used some of the LEP2
 results on fermion pair production in order to constrain the 
set of constants 
$\delta_{Z, s, \gamma}$ which fully describes
 the effects of general Universal
 New Physics.
In particular, in the same spirit of a previous study \cite{deltai}, 
we have considered the following ``non Bhabha'' observables: 
$\sigma_{\mu, \tau}$ (the cross section for $\mu$ and $\tau$ pair production), 
$A_{FB, \mu, \tau}$ (the related forward-backward asymmetries) and
 $\sigma_5$ (the cross section for production of quark pairs, for the
 five light flavours accessible at LEP energies). In addition to these
 observables we have included the unpolarized differential
 Bhabha cross section, measured in intervals of the cosine of the polar
 angle of the scattered electron. 
For the muon and tau cross sections and asymmetries and for the hadronic
 cross section the combinations of preliminary results of the four LEP
 experiments with the data collected at center of mass energies of 183~GeV
 and 189~GeV have been used \cite{LEP2comb}. Although based on preliminary
 results, the combined measurements allow to take advantage of the whole 
 data sample produced at LEP2 and therefore to benefit of the reduced 
 statistical error and of the proper treatment of the various sources of 
 experimental systematic uncertainty. 
 A measurement of the differential cross
 section for Bhabha scattering with acollinearity smaller that 10$^\circ$ has
 been recently
performed with a data sample of approximately 180~$pb^{-1}$ at the
 center of mass energy of 189~GeV \cite{OPALbb}. The differential cross
 section is measured in nine uniform intervals of the polar angle of the scattered
 electron, $\cos\theta_{e^-}$, in the range (-0.9, 0.9). The precision of
 the measurement, which is limited primarely by the statistical uncertainty,
 reaches the level of 1\% in the interval of most forward scattering
 angles. 

 This measurement, together with the combined results on muon, 
 tau and hadronic observables, has been compared to the Standard Model 
 prediction corresponding to the experimental signal definition. The
 deviations of the measurement with respect to the Standard Model
 expectations have then been used to measure and constrain the 
 parameters $\delta_Z$, $\delta_s$ and $\delta_\gamma$ with a $\chi^2$ fit. 
 In the fit procedure the uncertainty on the reference Standard
 Model prediction itself must be taken into account. The
 theoretical uncertainties on the Standard Model predictions for
 fermion pair production at LEP2 energies are mainly related to the
 estimate of the large QED corrections. In the case of $\mu^+\mu^-$,
 $\tau^+\tau^-$ and $q\bar{q}$ production, the differences between 
 predictions of several semianalytic or Monte Carlo calculations 
 \cite{ZF,TZ0,KK} for cross sections and asymmetries 
are smaller than 1\% and, therefore, they are negligible with respect
to the experimental error. 
 In the Bhabha scattering process the different programs
 \cite{TZ0,BHWIDE,ALIBABA} providing the Standard Model predictions compare
 each other at the level of 2\% in the experimental
acceptance. 
Therefore, in our study, we have assigned 
 a 2\% uncertainty to the reference Standard Model prediction for the
 Bhabha differential cross section. This uncertainty reflects into an error
 larger than the experimental one in the region of forward scattering,
 which, as will be discussed in the following, is the most sensitive to New
 Physics effects in $\delta_\gamma$. 

The results of the analysis are shown in Tab.~(\ref{tab:1}). As one sees, the
addition of Bhabha scattering improves,
although not spectacularly, the bound on $\delta_\gamma$ which is 
constrained by the data in the forward scattering angle region, 
where the Bhabha cross section is dominated by the $t$-channel photon
exchange contribution. It is interesting to take in mind that , at present,
the sensitivity to NP effects in the forward Bhabha cross section is
spoiled by the theoretical uncertainty on the Standard Model theoretical
prediction, which dominates over the experimental error. 
 Should this error
be reduced, the role of this measurements would certainly be more relevant.

To give a more quantitative meaning to the latter claim, we have 
simulated a forthcoming measurement at 200 GeV with an overall $400$ 
$pb^{-1}$ luminosity for each experiment, and repeated the previous 
analysis adding these future data to those available at $183$, $189$ GeV.
For consistency, we have assumed in all the three sets of data a central
value coincident with the SM prediction. The errors are
those available at $183$ and $189$ GeV. At $200$ GeV
we considered two scenarios, one with purely statistical errors and one
with their combination with a 2\% theoretical error on Bhabha
scattering. 
The difference 
between the two cases affects mainly the very forward scattering cone
and therefore $\delta_\gamma$.
The
results of this second analysis are shown in Tab.~(\ref{tab:2}).
As one sees, in agreement with the qualitative expectations, the role
of Bhabha scattering is now definitely more relevant in the 
determination of the bound for the photonic parameter $\delta_\gamma$. 

The
same two analyses have been performed for the two models involving
contact terms and extra dimensions. The results are presented in
Tables III,IV. For what concerns the contact terms, one should first
note that the values of the bounds obtained from $e^+e^-\to
\mu^+\mu^-,~q\bar q$ (without Bhabha) depend strongly on the chirality
structure. This comes from the interference of the contact amplitude
with the standard $\gamma,~Z$ exchange amplitude, which is larger
when the chirality structures of both amplitudes are close
to each other. In particular the $VA$ bound is found very low because
the standard $VA$ amplitude is depressed by the small $Ze^+e^-$
vector coupling. One then sees that the effect of the Bhabha process
on these bounds is generally modest. An opposite situation appears
for the case of extra dimensions, where the bulk of the effect 
is provided
by the ``forward'' data. As already mentioned in Section IIIB(4), the
largest effect on the differential cross section comes from the
interference of the standard photon exchange with the extra dimension
term both in the $t$-channel, followed by the mixed ($s$-channel)-
($t$-channel) one, the ($s$-channel)-($s$-channel) contribution
being much smaller.
Once again, this is more clearly visible in the
analysis that uses the future data at 200 GeV, in the case one assumes
no theoretical error in the the Bhabha component.

\begin{figure}
\caption{
Two dimensional projections of the 95\% C.L. allowed region
in the  $\delta_{s,Z,\gamma}$  space from a global fit of LEP2 
results on two fermion production.
The observables included in the fit are $\sigma_{\mu, \tau}$, 
$A_{FB, \mu, \tau}$, $\sigma_5$ measured at 183 and 189 GeV and 
the Bhabha unpolarized differential cross section 
(in nine intervals of the cosine of the scattering angle)
at 189 GeV. The inner ellipses are the projection of the intersection 
of the 3 dimensional ellipse and the AGC (dashed line) or TC (dotted line)
constraints. The small cross marks the axes origin, corresponding to the 
Standard Model case.
}
\begin{center}
\epsfig{file=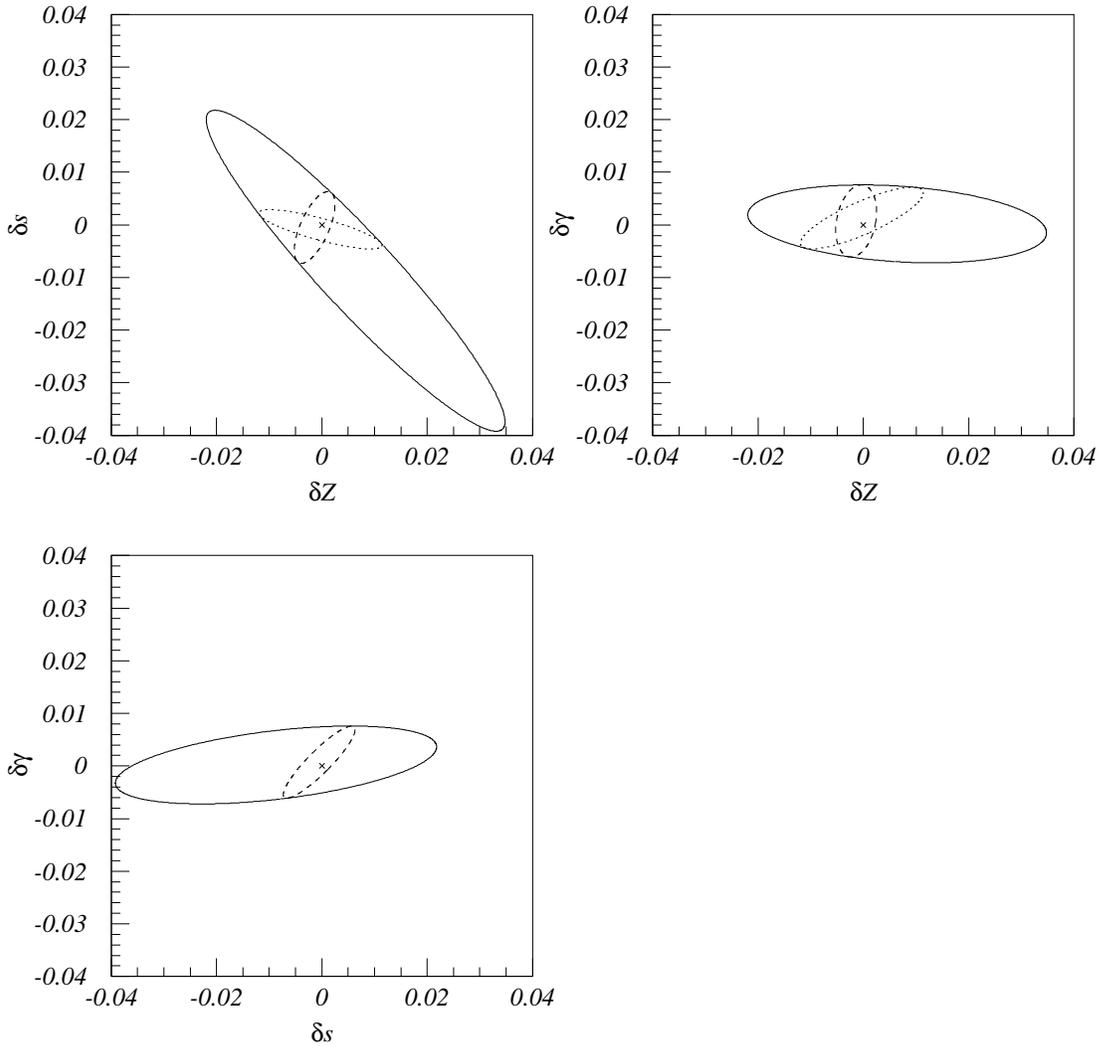,width=16cm}
\end{center}
\label{fig:1}
\end{figure}

\begin{figure}
\caption{
Two dimensional projections of the 95\% C.L. allowed region
in the  $\delta_{s,Z,\gamma}$  space from a global fit of LEP2 
results on two fermion production.
The observables included in the fit are $\sigma_{\mu, \tau}$, 
$A_{FB, \mu, \tau}$, $\sigma_5$ measured at 183, 189, 200 GeV and 
the 9 angular measurements of the Bhabha unpolarized cross section 
at 189 and 200 GeV. 
We always assumed the experimental measurements to be coincident with the
Standard Model predictions. About the errors, we took the available 
actual experimental errors for measurements at 183 and 189 GeV (like in 
the previous figure) and a purely statistical error 
at 200 GeV under the assumption of an integrated luminosity of 
400 $pb^{-1}$ for each of the 4 LEP2 experiments.
The inner ellipses are the projection of the intersection 
of the 3 dimensional ellipse and the AGC (dashed line) or TC (dotted line)
constraints.
}
\begin{center}
\epsfig{file=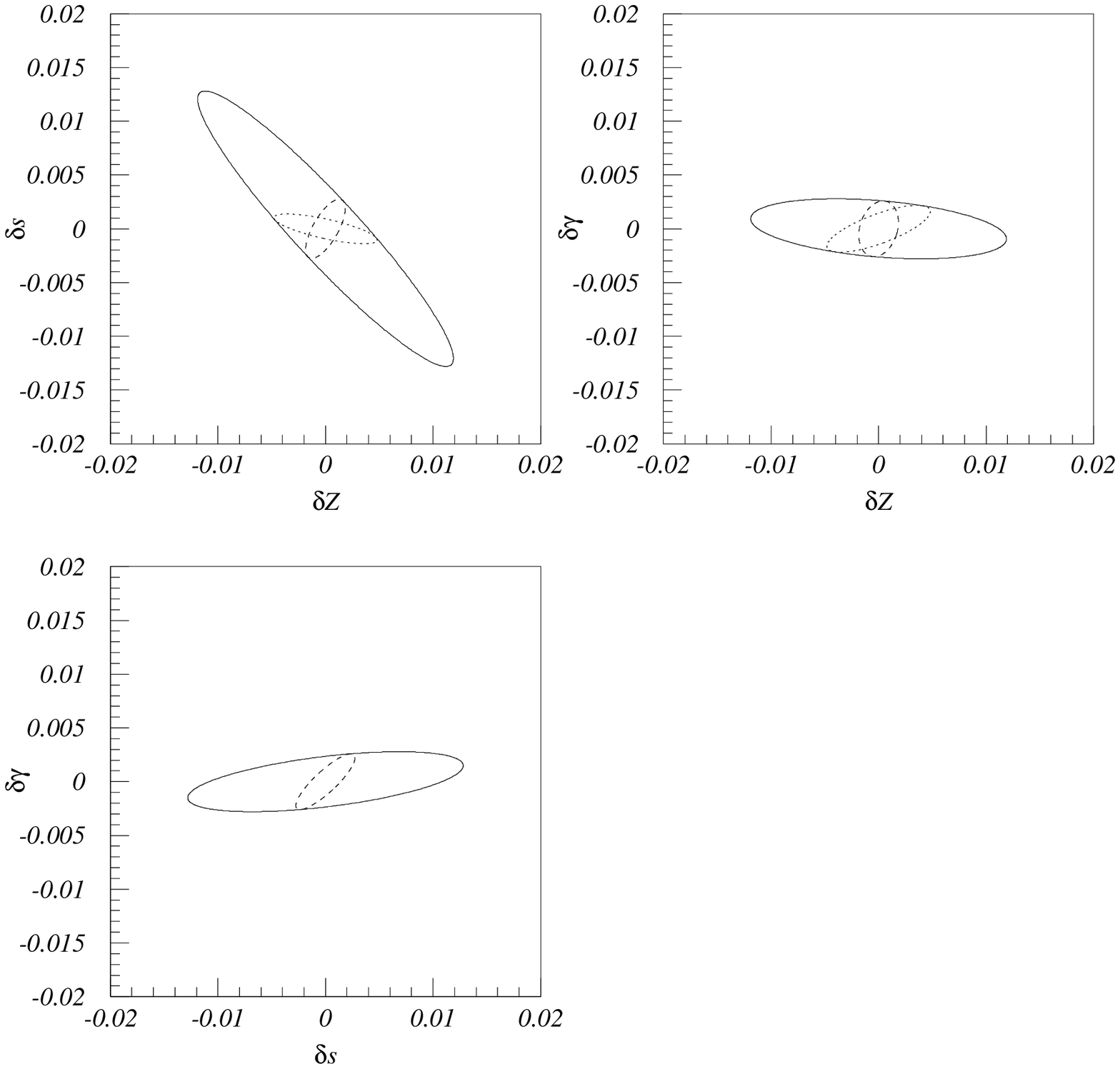,width=16cm}
\end{center}
\label{fig:2}
\end{figure}

All the numerical results exhibited in Tabs.~(\ref{tab:1},\ref{tab:2})
can be represented graphically. 
We have shown in Figs.~(\ref{fig:1},\ref{fig:2})
the planar ellipses that are obtained by projecting onto the three planes 
$(\delta_s, \delta_\gamma)$,
$(\delta_Z, \delta_\gamma)$ and
$(\delta_Z, \delta_s)$ the 95 \%  C. L. allowed three dimensional 
region resulting from a global fit of all data in terms of the three
parameters $\delta_{Z, s, \gamma}$. For completeness, we also show
the results for the two representative AGC and TC models. As a very 
preliminary comment concerning the latter cases we can notice that, 
although at a rather qualitative level, LEP2 data apparently do not 
particularly support the considered TC theoretical proposal that would 
require $\delta_s>0$.

\subsection{LC analyses}

This analysis has been performed in a spirit that is very
similar to that used for the future 200 GeV\ LEP2 analyses. In other
words, we have assumed a set of measurements at $\sqrt{q^2}=500$ GeV
whose central values agree with the SM predictions, and postulated 
a purely statistical error corresponding to a high luminosity of  
500 $fb^{-1}$. 
We have added to the previous LEP2 ``non Bhabha''
observables the longitudinal polarization asymmetry $A_{LR}$ for
lepton production. A discussion of the important role of this 
observable can be found in~\cite{alr}. 
Of course, in principle other measurements e.g. for
final $b$ or $t$ quarks could be used. We have also assumed 9
angular Bhabha measurements  for all the three different observables
($\sigma$, $\sigma^1$, $\sigma^P$) defined in Section~(2). More precisely,
in each bin $[\theta_{\min}, \theta_{max}]$ we have considered 
\bq
\sigma_{[\theta_{min}, \theta_{max}]} = 
\int_{\cos\theta_{max}}^{\cos\theta_{min}}
 \frac{d\sigma}{d\cos\theta}\ d\cos\theta
\eq
\bq
A_{LR, [\theta_{min}, \theta_{max}]} = \frac{1}{\sigma_{[\theta_{min}, \theta_{max}]}}
\int_{\cos\theta_{max}}^{\cos\theta_{min}}
 \frac{d\sigma^P}{d\cos\theta}\ d\cos\theta, \quad
\eq
\bq
A_{||, [\theta_{min}, \theta_{max}]} = \frac{1}{\sigma_{[\theta_{min}, \theta_{max}]}}
\int_{\cos\theta_{max}}^{\cos\theta_{min}}
 \frac{d\sigma^2}{d\cos\theta}\ d\cos\theta, \quad
\eq

Tables~(\ref{tab:5}, \ref{tab:6}) contain the numerical
results for the universal and not universal models. As a general feature, 
one notices that in the universal case the limits on \underline{all}
the three parameters $\delta_{Z,s,\gamma}$ are substantially 
(a factor of two) improved by the use of Bhabha observables. The 
interesting feature is that, in each case, different Bhabha observables
play the crucial role. In fact, $\delta_Z$ is mostly affected by $A_{||}$
(in both angular directions), $\delta_s$ (as one expects) by $A_{LR}$ ( in 
the forward cone), and $\delta_\gamma$ by the unpolarized $\sigma$
 (again, for very small angles). In the considered non universal cases, the 
effect is, again, not spectacular (although not negligible)
for the contact terms. Quite on the contrary, there would be a large
(a factor two) effect in the case of extra dimensions, mostly due to the
unpolarized cross section at small angles. For this specific model of 
new physics Bhabha scattering seems therefore to represent
a fundamental experimental measurement. This statement is well in agreement
with the results of a recent numerical analysis of LEP2 data~\cite{bou}.

\newpage
\section{Conclusion}

In conclusion, we have shown that, for a class of theoretical models of new
physics that is certainly not empty, the generalization of the $Z$-peak
subtracted approach to the case of Bhabha scattering can be simply
performed, leading in general to improvements of the information that 
might be obtained. As a general statement, Bhabha scattering appears to
be always relevant; for models of universal type, polarized and
unpolarized observables play a crucial role in the determination of the
bounds for the different parameters; in other non universal interesting
cases, like in particular that of extra dimensions, unpolarized Bhabha
observables appear to play a fundamental role.

\newpage

{\large \bf Appendix A: General form of the polarized Bhabha
scattering cross section.}\\

\bq
{d\sigma\over dcos\theta}=(1-PP'){d\sigma^{1}\over dcos\theta}
+(1+PP'){d\sigma^{2}\over dcos\theta}
+(P'-P){d\sigma^{P}\over dcos\theta}
\eq

with
\bqa
&&{d\sigma^{N1}\over dcos\theta}={1\over16\pi q^2}\{{t^2+u^2\over
q^4}(g^{(\gamma)}_{Ve})^4 + {2(q^2-M^2_Z)\over
q^2[(q^2-M^2_Z)^2+M^2_Z\Gamma^2_Z]}[(t^2+u^2)(g^{(\gamma)}_{Ve}
g^{(Z)}_{Ve})^2+(u^2-t^2)(g^{(\gamma)}_{Ve}
g^{(Z)}_{Ae})^2]\nonumber\\
&&+{1\over[(q^2-M^2_Z)^2+M^2_Z\Gamma^2_Z]}[(t^2+u^2)[(g^{(Z)}_{Ve})^2
+(g^{(Z)}_{Ae})^2]+4(u^2-t^2)(g^{(Z)}_{Ve}g^{(Z)}_{Ae})^2]
+{2u^2\over q^2t}(g^{(\gamma)}_{Ve}\gb^{(\gamma)}_{Ve})^2\nonumber\\
&&+{2u^2\over
q^2(t-M^2_Z)}(g^{(\gamma)}_{Ve})^2[(\gb^{(Z)}_{Ve})^2
+(\gb^{(Z)}_{Ae})^2]+{2u^2(q^2-M^2_Z)\over t
[(q^2-M^2_Z)^2+M^2_Z\Gamma^2_Z]}(\gb^{(\gamma)}_{Ve})^2
[(g^{(Z)}_{Ve})^2+(g^{(Z)}_{Ae})^2]\nonumber\\
&&+{2u^2(q^2-M^2_Z)\over (t-M^2_Z)
[(q^2-M^2_Z)^2+M^2_Z\Gamma^2_Z]}([(g^{(Z)}_{Ve})^2+(g^{(Z)}_{Ae})^2]
[(\gb^{(Z)}_{Ve})^2+(\gb^{(Z)}_{Ae})^2]+4g^{(Z)}_{Ve} g^{(Z)}_{Ae}
\gb^{(Z)}_{Ve}\gb^{(Z)}_{Ae})\nonumber\\
&& +{u^2\over
t^2}(\gb^{(\gamma)}_{Ve})^4+{2u^2\over t(t-M^2_Z)}
(\gb^{(\gamma)}_{Ve})^2[(\gb^{(Z)}_{Ve})^2+(\gb^{(Z)}_{Ae})^2]
+{u^2\over(t-M^2_Z)^2}[((\gb^{(Z)}_{Ve})^2+(\gb^{(Z)}_{Ae})^2)^2
+4 (\bar g^{(Z)}_{Ve}\bar g^{(Z)}_{Ae})^2]\}\nonumber\\
\label{N1}\eqa

\bqa
&&{d\sigma^{2}\over dcos\theta}={1\over16\pi q^2}\{{q^4\over
t^2}(\bar g^{(\gamma)}_{Ve})^4 + {2q^4\over
t(t-M^2_Z)}(\gb^{(\gamma)}_{Ve})^2[(g^{(Z)}_{Ve})^2-
(g^{(Z)}_{Ae})^2]\nonumber\\
&&+{q^4\over(t-M^2_Z)^2}([(\gb^{(Z)}_{Ve})^2
-(\gb^{(Z)}_{Ae})^2]^2)\}
\label{N2}\eqa

\bqa
&&{d\sigma^{P}\over dcos\theta}={u^2\over4\pi q^2}\{
{1\over[(q^2-M^2_Z)^2+M^2_Z\Gamma^2_Z]}({q^2-M^2_Z\over
q^2}(g^{(\gamma)}_{Ve})^2(g^{(Z)}_{Ve}
g^{(Z)}_{Ae})+(g^{(Z)}_{Ve}
g^{(Z)}_{Ae})[(g^{(Z)}_{Ve})^2
+(g^{(Z)}_{Ae})^2])\nonumber\\
&&+{1\over
q^2(t-M^2_Z)}(g^{(\gamma)}_{Ve})^2(\gb^{(Z)}_{Ve}
\bar g^{(Z)}_{Ae})+{q^2-M^2_Z)\over t[(q^2-M^2_Z)^2+M^2_Z\Gamma^2_Z]}
(\gb^{(\gamma)}_{Ve})^2(g^{(Z)}_{Ve}
g^{(Z)}_{Ae})\nonumber\\
&&+{q^2-M^2_Z\over(t-M^2_Z)[(q^2-M^2_Z)^2+M^2_Z\Gamma^2_Z]}
((g^{(Z)}_{Ve}
g^{(Z)}_{Ae})[(\gb^{(Z)}_{Ve})^2+(\gb^{(Z)}_{Ae})^2]+
(\gb^{(Z)}_{Ve}
\gb^{(Z)}_{Ae})[(g^{(Z)}_{Ve})^2+(g^{(Z)}_{Ae})^2])\nonumber\\
&&
+{1\over t(t-M^2_Z)}(\gb^{(\gamma)}_{Ve})^2(\gb^{(Z)}_{Ve}
\gb^{(Z)}_{Ae})+{1\over(t-M^2_Z)^2}(\gb^{(Z)}_{Ve}
\gb^{(Z)}_{Ae}
[(\gb^{(Z)}_{Ve})^2+(\gb^{(Z)}_{Ae})^2]\}
\label{P}\eqa

\newpage

\begin{table}
\caption{{\bf LEP2}.
95\% C.L. bounds on $\delta_{s,Z,\gamma}$ resulting from 
a global fit of combined LEP2 data.
In the first column, 
the observables included in the fit are $\sigma_{\mu, \tau}$, 
$A_{FB, \mu, \tau}$, $\sigma_5$ measured at 183 and 189 GeV.
In the second column, we add the 
Bhabha unpolarized differential cross section ( in nine intervals of
the cosine of the scattering angle)
at 189 GeV. The last two columns show the improvement of the bounds
when only the forward (backward) Bhabha scattering measurements are 
included.
}

\begin{center}
\begin{tabular}{c|cccc}
               & without Bhabha & with all Bhabha  & forward & backward 	\\
\hline
$\delta_Z$      & -0.001 $\pm$ 0.031 &	0.0064 $\pm$ 0.028 & 0.006 $\pm$ 0.03  & 0.0011 $\pm$ 0.029	\\
$\delta_s$      & -0.004  $\pm$ 0.032 & -0.0087 $\pm$ 0.031 & -0.0084 $\pm$ 0.032 & -0.0057 $\pm$ 0.031	\\
$\delta_\gamma$ & -0.0022  $\pm$ 0.0083  & 0.00019 $\pm$ 0.0074& 0.00014 $\pm$ 0.0075 & -0.0019 $\pm$ 0.0081
\end{tabular}
\end{center}
\label{tab:1} 
\end{table}

\begin{table}
\caption{{\bf LEP2}. 
95\% C.L. bounds on $\delta_{s,Z,\gamma}$ resulting from 
a global fit of the simulated forthcoming combined LEP2 data.
The observables included in the fit are $\sigma_{\mu, \tau}$, 
$A_{FB, \mu, \tau}$, $\sigma_5$ at 183, 189, 200 GeV and, in the
second column, also the 9 angular measurements of the 
Bhabha unpolarized cross section at 189 and 200 GeV. 
For all the observables we assume that the measurement are coincident
with the Standard Model predictions. The errors are the experimental
ones at 183 and 189 GeV. At 200 GeV, the errors are the statistical
ones associated to a
400 $pb^{-1}$ integrated luminosity per experiment with or without 
(second and third columns) a $2\%$ theoretical error on Bhabha scattering.
}

\begin{center}
\begin{tabular}{c|ccc}
               & without Bhabha & with all Bhabha and 2\% th. err & with all Bhabha \\
\hline
$\delta_Z$      & 0.014  & 0.012 & 0.012\\
$\delta_s$      & 0.015  & 0.013 & 0.013\\
$\delta_\gamma$ & 0.0038 & 0.0034 & 0.0028
\end{tabular}
\end{center}
\label{tab:2}
\end{table}

\begin{table}
\caption{{\bf LEP2}. 
Bounds on non universal New Physics scale at 95\% C.L. 
resulting from 
a global fit of the present combined LEP2 data.
We consider
the models of Contact Interactions (in the eight cases LL, RR, LR, RL, VV, AA,
AV, VA)
and Extra Dimensions discussed in the paper. The data sets used for the 
fit are the same as in Tab.~(\ref{tab:1}).
Since the central value for $1/\Lambda$ is not zero, we show the two 
different bounds for $|\Lambda|$ that are obtained in the two cases
$\Lambda<0$ and $\Lambda>0$ respectively.}

\begin{center}
\begin{tabular}{c|cccc}
                 & without Bhabha & with all Bhabha & forward & backward  \\
\hline
$\Lambda_{LL}$ &10-9.9  &11-9.2 &11-9.2 &10-9.8 \\
$\Lambda_{RR}$ &7.7-12  &8.7-10 &8.7-10 &7.8-12 \\
$\Lambda_{LR}$ &6.5-9.2 &16-7.8 &14-7.3 &8.2-9.2 \\
$\Lambda_{RL}$ &7.2-15  &12-9.7 &11-9.5 &8.3-13 \\
$\Lambda_{VV}$ &13-20   &17-16  &16-16  &13-20 \\
$\Lambda_{AA}$ &16-13   &14-14  &14-14  &15-13 \\
$\Lambda_{AV}$ &17-8.7  &17-8.7 &17-8.7 &17-8.7 \\
$\Lambda_{VA}$ &4-3.3   &4.2-3.2&4.2-3.2&4-3.3 \\
\hline
$\Lambda_{ED}$ &0.69-0.75& 0.82-2.2& 0.8-1.9& 0.77-0.9 
\end{tabular}
\end{center}
\label{tab:3}
\end{table}

\begin{table}
\caption{{\bf LEP2}. 
Bounds on non universal New Physics scale at 95\% C.L. 
resulting from 
a global fit of the simulated forthcoming combined final 
LEP2 data. Data sets are chosen as in Tab.~(\ref{tab:2}).}

\begin{center}
\begin{tabular}{c|ccc}
                 & without Bhabha & with all Bhabha and 2\% th. err & with all Bhabha \\
\hline
$\Lambda_{LL}$ &15      &15     &16 \\
$\Lambda_{RR}$ &13      &14     &15 \\
$\Lambda_{LR}$ &11      &16     &18 \\
$\Lambda_{RL}$ &13      &17     &18 \\
$\Lambda_{VV}$ &22      &24     &27 \\
$\Lambda_{AA}$ &21      &21     &22 \\
$\Lambda_{AV}$ &16      &16     &16 \\
$\Lambda_{VA}$ &5.2     &5.2    &5.3 \\
\hline
$\Lambda_{ED}$ & 0.89 & 1.2 &1.4 
\end{tabular}
\end{center}
\label{tab:4}
\end{table}

\begin{table}
\caption{{\bf Linear Collider}. 
95\% C.L. bounds on $\delta_{s,Z,\gamma}$ resulting from 
a global fit of LC data assuming that data are taken at 500 GeV
with an integrated high luminosity 500 $fb^{-1}$.
The ``non-Bhabha'' observables included in the fit are $\sigma_{l}$, 
$A_{FB, l}$, $A_{LR, l}$ and $\sigma_5$, where $l$ stands for lepton.
The ``Bhabha'' observables are the unpolarized angular distribution
and the two asymmetries $A_{LR}$ and $A_{||}$ defined in the paper.
Each of these three observables is assumed to be measured in the 
same 9 bins as in the LEP2 analysis.
The central values are assumed to coincide with the Standard Model 
predictions and the errors are purely statistical.
The first column is obtained without including the Bhabha observables 
in the fit.
The second column is with all the Bhabha observables. The next three columns
are obtained including in the fit only one of the three $\sigma$, $A_{LR}$
and $A_{||}$. Finally, the last two columns are obtained by using the 
three observables, but only in the forward (backward) cone.
}

\begin{center}
\begin{tabular}{c|ccccccc}
                      & without Bhabha & with all Bhabha  & $\sigma$ & $A_{LR}$ & $A_{||}$ & forw. & back. \\
\hline
$10^4\ \delta_Z$      & 2.4  & 1.3    & 1.9      & 2.1      & 1.7      & 1.6  & 1.7 \\
$10^4\ \delta_s$      & 1.4  & 0.8    & 1.3      & 0.85     & 1.3      & 0.82 & 1.3 \\
$10^4\ \delta_\gamma$ & 1.1  & 0.56   & 0.62     & 1.1      & 0.9      & 0.65 & 0.82 
\end{tabular}
\end{center}
\label{tab:5}
\end{table}

\begin{table}
\caption{{\bf Linear Collider}. 
95\% C.L. bounds on non universal New Physics scales
for the considered 
Contact Interactions and Extra Dimensions models.
Data sets and columns meaning are as in Tab.~(\ref{tab:5}).
Units are TeV.}

\begin{center}
\begin{tabular}{c|ccccccc}
    & without Bhabha & with all Bhabha & $\sigma$ & $A_{LR}$ & $A_{||}$ & forw. & 
back. \\
\hline
$\Lambda_{LL}$ & 85 & 94 & 89 & 89 & 87 & 93 & 85\\
$\Lambda_{RR}$ & 84 & 92 & 87 & 88 & 85 & 92 & 84\\
$\Lambda_{LR}$ & 66 & 120 & 120 & 66 & 87 & 110 & 110\\
$\Lambda_{RL}$ & 81 & 130 & 120 & 81 & 94 & 110 & 110\\
$\Lambda_{VV}$ & 120 & 150 & 150 & 120 & 120 & 150 & 140\\
$\Lambda_{AA}$ & 110 & 140 & 130 & 110 & 120 & 120 & 130\\
$\Lambda_{AV}$ & 130 & 130 & 130 & 130 & 130 & 130 & 130\\
$\Lambda_{VA}$ & 71 & 90 & 71 & 90 & 71 & 90 & 71\\
\hline
$\Lambda_{ED}$ & 3.3 & 5.7 & 5.7 & 3.3 & 3.6 & 5.6 & 4.5 
\end{tabular}
\end{center}
\label{tab:6}
\end{table}

\newpage


\begin{thebibliography}{99}


\bibitem{LEPcombi} The LEP Collaborations ALEPH, DELPHI, L3, Opal, the
LEP Electroweak Working Group and the SLD Heavy Flavour and Electroweak
Groups, CERN-EP/99-15.

\bibitem{Zsub} 
F.M.~Renard and C.~Verzegnassi, 
Phys. Rev. {\bf D52}, 1369 (1995), 
Phys. Rev. {\bf D53}, 1290 (1996).

\bibitem{applic}  
F.M. Renard and C. Verzegnassi
Phys. Rev. {\bf D55}, 4370 (1997);
%
M. Beccaria, G. Montagna, O. Nicrosini, F. Piccinini,
F.M. Renard, C. Verzegnassi,
Phys. Rev. {\bf D58}, 093014 (1998);
%
M. Beccaria, P. Ciafaloni, D. Comelli, F. Renard, C. Verzegnassi,
EuroPhys. J. {\bf C10}, 331 (1999).

\bibitem{agczsub} 
A. Blondel, F. M. Renard, L. Trentadue and
C. Verzegnassi, 
Phys. Rev. {\bf D54}, 5567 (1996);
%
M. Beccaria, F.M. Renard, S. Spagnolo and C. Verzegnassi, 
Phys. Lett. {\bf B448}, 129 (1999).

\bibitem{tc}
R. S. Chivukula, F. M. Renard and C. Verzegnassi, 
Phys. Rev. {\bf D57}, 2760 (1998).

\bibitem{deltai}  
M. Beccaria, F.M. Renard, S. Spagnolo and C. Verzegnassi, 
hep-ph/9911372, to appear in Phys. Lett. B.

\bibitem{Peskin} 
M.E. Peskin, T. Takeuchi, 
Phys. Rev. Lett. {\bf 65}, 964 (1990).

\bibitem{AB}  
G. Altarelli and R. Barbieri, 
Phys. Lett. {\bf B253}, 161 (1991).

\bibitem{log}  
M. Beccaria, P. Ciafaloni, D. Comelli, F.M. Renard and C. Verzegnassi,
Phys. Rev. {\bf D61},073005(2000).

\bibitem{agc} 
K. Hagiwara, S. Ishihara, R. Szalapski and D. Zeppenfeld, 
Phys. Rev. {\bf D48}, 2182 (1993).


\bibitem{contact} 
E. Eichten, K. Lane, M. Peskin, 
Phys. Rev. Lett. {\bf 50}, 811 (1983).

\bibitem{Schrempp}
B. Schrempp, F. Schrempp, N. Wermes and D. Zeppenfeld, Nucl. Phys.
{\bf B296},1 (1988). 

\bibitem{graviton} 
J. Hewett, 
Phys. Rev. Lett. {\bf 82},  4765 (1999);
%
T. Rizzo, 
Phys. Rev. {\bf D59}, 115010 (1999).
 
\bibitem{LEP2comb} LEP EWWG $f\bar{f}$ Subgroup, ``Combination of the LEP2 
   $f\bar{f}$ Results'', LEP2FF/99-01, ALEPH 99-082 PHYSIC 99-030, DELPHI 99-143 PHYS 829, 
   L3 Note 2443, OPAL TN616; 
   LEP EWWG $f\bar{f}$ Subgroup: http://www.cern.ch/LEPEWWG/lep2/.

\bibitem{OPALbb} The OPAL Collaboration, G. Abbiendi \etal\ ``Test of the
  Standard Model and Constraints on New Physics from Measurements of
  Fermion Pair Production at 189 GeV at LEP'', CERN-EP/99-097, July 1999;

\bibitem{ZF} ZFITTER; 
D.~Bardin et~al., Phys. Lett. {\bf B255}, 290 (1991); 
D.~Bardin et~al., Nucl. Phys. {\bf B351}, 1 (1991); 
D.~Bardin et~al., Z.Phys. {\bf C44}, 493 (1989).

\bibitem{TZ0} TOPAZ0; 
G.~Montagna, O.~Nicrosini, G.~Passarino, F.~Piccinini and R.Pittau,
Comp. Phys. Comm. {\bf 76}, 328 (1993).

\bibitem{KK} KK2f; 
S.~Jadach, B.F.L.~Ward and Z.~W\c{a}s, 
Phys. Lett. {\bf B449}, 97 (1999).

\bibitem{BHWIDE} BHWIDE; 
S.~Jadach, W.~Placzek, B.F.L.~Ward, 
Phys. Lett. {\bf B390}, 298 (1997).

\bibitem{ALIBABA} ALIBABA; 
W.~Beenakker~et~al., 
Nucl. Phys. {\bf B349}, 323  (1991).



%













\bibitem{alr} 
F. M. Renard and C. Verzegnassi, 
Phys. Rev. {\bf D55}, 4370 (1997); 
%
M. Beccaria, F. M. Renard, S. Spagnolo and C. Verzegnassi, 
``New Physics Effects from $e^+e^-\to\bar{f} f$ at a
Linear Collider: the role of $A_{LR, \mu}$, 
DESY/ECFA note LC-TH-1999-016.

\bibitem{bou} D. Bourilkov, JHEP 9908, 006 (1999).


\end{thebibliography}
\end{document}